\documentclass{aastex}
\usepackage{spr-astr-addons}
\usepackage{url}

\RequirePackage{color}

\begin{document}

\title{Magnetized Kelvin-Helmholtz instability in the presence of a radiation field}
\slugcomment{Not to appear in Nonlearned J., 45.}
%% Running heads
\shorttitle{Short article title}
\shortauthors{Autors et al.}

\author{Mohsen Shadmehri\altaffilmark{1}}
\affil{School of Physics, Faculty of Science, Golestan University, Gorgan, Iran}
\and
\author{Zahra Enayati\altaffilmark{2}} \and \author{Mahdi Khajavi\altaffilmark{2}}
\affil{School of Physics, Faculty of Science, Ferdowsi University, Mashhad, Iran}
%\email{\emaila}

%\altaffiltext{1}{School of Physics, Faculty of Science, Ferdowsi University, Mashhad, Iran.}
\altaffiltext{1}{m.shadmehri@gu.ac.ir}
%\altaffiltext{3}{Third Alternate Affilation.}

\begin{abstract}
 The purpose of this study is to analyze the dynamical role of a radiation field on the growth rate of the unstable Kelvin - Helmholtz (KH) perturbations. As a first step toward this purpose, the analyze is done in a general way, irrespective of applying the model to a specific astronomical system. The transition zone between the two layers of the fluid is ignored. Then, we  perform a linear analysis and by imposing suitable boundary conditions and considering a radiation field, we obtain appropriate  dispersion relation. Unstable modes are studied by solving the dispersion equation numerically, and then growth rates of them are obtained. By analyzing our dispersion relation, we show that for a wide range of the input parameters, the radiation field has a destabilizing effect on KH instability. In eruptions of the galaxies or supermassive stars, the radiation field is dynamically important and because of the enhanced KH growth rates in the presence of the radiation; these eruptions can inject more momentum and energy into their environment and excite more turbulent motions.
\end{abstract}


%\section*{}
%\label{sec:intro}

\section{Introduction}

Kelvin-Helmholtz (KH) instability occurs when different layers of a fluid are in relative motion. This instability occurs in many astrophysical systems and has been widely studied in relation to the solar wind \citep{amer,betta,hasegawa},  pulsar winds \citep{bucc} and thermal flares \citep{venter}. The KH instability can also drive mixing and turbulence and is thus relevant in protoplanetary discs \citep{Johan,gomez}, accretion discs and magnetospheres \citep{LiNarayan} and other jets and outflows \citep{baty}.

\citet{chandra} studied  KH instability in the linear regime. Using perturbation analysis, he derived the conditions and the timescales required for the growth of this instability. Also examined were the effects of other parameters such as radiative cooling and magnetic fields, and their ability to stabilize the development of these non-linear structures. More recently,  KH instability has been investigated by means of numerical simulations, in both the relativistic \citep{Perucho2004,Perucho,bucc} and non-relativistic limits \citep{rossi, downes98, micono98, hana98, micono, michi}. One very important factor in all KH simulations is the role played by the magnetic field. Magnetohydrodynamics is required for all numerical simulations in order to accurately represent the presence of this field \citep{Viallet,Jeong,Min,shadmehri8, shadmehri10}.

But radiation may have dynamical effects in some of the astronomical objects. For example, formation of the stars takes place in the interstellar medium (ISM) where a wide variety of the structures and the physical conditions are observed. There is a growing interest toward understanding properties of the ISM, in particular regarding to the dynamical role of the radiation in the gravitational instability and possibly massive star formation \citep{Vranjes90,Vranjes990}.

Dynamical role of the radiation field in the context of the accretion flows has also been studied by many authors. At the larger  scales, we have radiation dominated discs. In particular, the very inner parts of the accretion discs around black holes are radiation pressure dominated, though there are many uncertainty about their physical state. In the standard model of accretion discs   \citep{Pringle,Shakura}, the viscous stress is assumed to scale with the radiation pressure, but the disc itself is subject to the thermal and the viscous instabilities \citep{Sunyaev, Lightman}. In addition to these  instabilities, dynamical instabilities also exist. First and foremost, magnetorotational instability (MRI) \citep{Hawley} is effective in weakly magnetized accretion discs. Gammie (1998) has suggested that the overstable photon bubble modes discussed by \citeauthor{Arons}(1992) in the X-ray pulsars also exist in radiation pressure dominated accretion flows in general. These dynamical instabilities may all play a role at some level in the dynamics and thermodynamics of the radiation pressure dominated portion of accretion disks \citep{Socrates}.

According to the observations of the galaxies only a few percent of the available gas reservoir galaxies is converted into stars per local free-fall time \citep{Krumholz, Kennicutt}. In other words, the efficiency of star formation in the galaxies is low. Injection of energy and momentum into the interstellar medium (ISM) by stellar processes (feedback), may be responsible for the inefficiency of star formation in galaxies and for their self-regulation. Various models for feedback, effects of HII gas pressure, shocked stellar winds, protostellar jets have been proposed over recent years. Dust grains are hydrodynamically coupled to the gas, and then radiation pressure on dust can help stabilize the gas against its own self-gravity and may therefore be an important feedback mechanism \citep{Thompson5}. When a jet or outflow interacts with its ambient medium, not only mass but energy are exchanged via the induced turbulence. One of the mechanisms that may initiate such a turbulence is KH instability. It has been proposed that winds may launch via a radiation-driven mechanisms \citep{Sharma, Murray}. Considering the dynamical effect of radiation in such system, it would be interesting to study the resulting KH instability at its interface with the ambient medium.

Early analytical studies of KH instability generally neglect an imbalance of the cooling and heating mechanisms. But subsequent analytical studies generalized the standard linear approach of KH instability to include heating and cooling. However, dynamical role of the radiation field has not been studied analytically or numerically as far as we know. Considering the importance of the radiation field in some of the astronomical systems subject to the KH instability as we described above, our goal is to investigate  KH instability including the dynamical role of the radiation. In the next section, basic assumptions and equations are presented. In section 3, we will study KH instability in the linear regime using a perturbation analysis. Different configurations are considered and we will generalize our study to include not only the radiation field but the magnetic field.

\section{Basic Equations}
Our main equations are the standard MHD equations including a radiation field assuming that the gas and the radiation are in perfect radiative equilibrium \citep{Mihalas, shadmehri7}. Thus,
\begin{equation}\label{eq:cont}
\frac{\partial\rho}{\partial t} + \nabla \cdot ({\rho {\bf u}}) =0,
\end{equation}
\begin{equation}\label{eq:momen}
\frac{\partial {\bf u}}{\partial t} + {\bf u} \cdot \nabla {\bf u} = -\frac{1}{\rho} \nabla (p_{\rm g} + p_{\rm r}) + \frac{1}{4\pi} ( {\bf J} \times {\bf B}),
\end{equation}
\begin{equation}\label{eq:induction}
\frac{\partial {\bf B}}{\partial t} = \nabla \times ({\bf u}\times {\bf B}),
\end{equation}
\begin{equation}\label{eq:polar}
\nabla \cdot {\bf B}=0,
\end{equation}
\begin{equation}\label{eq:energy}
\frac{\partial E}{\partial t} + {\bf u} \cdot \nabla E + (E + P ) \nabla \cdot {\bf u} = -\nabla \cdot {\bf F},
\end{equation}
\begin{equation}
{\bf F} = -\frac{c}{3\kappa \rho}\nabla e_{\rm r},
\end{equation}
\begin{equation}
p_{\rm g}=\rho c_{\rm s}^{2}.
\end{equation}
where the subscripts r and g refer to the radiation and the gas components, respectively. Here, $\rho$ and ${\bf u}$ are the density and the velocity of the gas. Also, the gas and the radiation and the total pressures are denoted by $p_{\rm g}$ and $p_{\rm r}$ and $P = p_{\rm g}+p_{\rm r}$, respectively. Moreover, we introduce the total internal energy  $E = e_{\rm g}+e_{\rm r}$, where  $e_{\rm g}=p_{\rm g}/(\gamma-1)$ and $e_{\rm r}=3p_{\rm r}$ and $\gamma$ is the adiabatic index of the gas.  Also ${\bf F}$ is the radiative flux,
and $\kappa$ is the opacity, {\bf B} is magnetic field and {\bf J} is the current density, i.e. ${\bf J}=\nabla\times{\bf B}$. We also assume that
$\kappa$ is constant, and for simplicity, the diffusion of the radiation is neglected.
\section{Linear Perturbations}
For doing a linear analysis, one should first specify the unperturbed properties of the system. We suppose that the streaming takes place in the
$x-$direction with a velocity $U(z)$ as
\begin{eqnarray*}
U(z) = \left\{
\begin{array}{rl}
U_{\rm 0},  &\qquad    z > 0\\
-U_{\rm 0}, &\qquad    z < 0
\end{array} \right.
\end{eqnarray*}
where $U_0$ is constant and all the unperturbed physical quantities
are assumed spatially uniform  in both the upper and the lower layers and we assume the initial magnetic field is along the streaming, i.e. ${\bf B}= B_{0}\hat{i}$ where $B_{0}$ is constant. In our configuration, the transition layer at $z=0$ has a negligible thickness.
Now, we can linearize the basic MHD equations.

We perturb the physical variables as $\zeta (x,z,t) = \zeta_{0} + \zeta'(x,z,t)$, where $\zeta_{0}$ is the unperturbed quantity and $|\zeta'|\ll |\zeta_0 | $. We substitute the perturbed variables as
\begin{equation}
\zeta'(x,z,t) = \zeta(z) e^{i(\omega t+ k_{\rm x}x)},
\end{equation}
where $\omega$ is the growth rate and $k_{\rm x}$ is the wavenumber of the perturbations. Thus, for a magnetized streaming fluid the  basic equations (\ref{eq:cont}), (\ref{eq:momen}), (\ref{eq:induction}), (\ref{eq:polar}), (\ref{eq:energy})  are linearized as
\begin{equation} \label{eq:1}
i \varphi \rho' + \rho_{\rm 0}(i k_{\rm x}u'_{\rm x} + \frac{du'_{\rm z}}{dz})=0,
\end{equation}
\begin{equation}\label{eq:2}
i \varphi \rho_{\rm 0} u'_{\rm x} = -i k_{\rm x}(p'_{\rm g}+p'_{\rm r}),
\end{equation}
%
%\begin{equation}\label{eq:2a}
%i \varphi \rho_{\rm 0} u'_{\rm y} = \frac{B_{0}}{4\pi}ik_{x}B'_{y},
%\end{equation}
%
\begin{equation}\label{eq:3}
i \varphi \rho_{\rm 0} u'_{\rm z} = -(\frac{dp'_{\rm g}}{dz}+\frac{dp'_{\rm r}}{dz})-\frac{B_{0}}{4\pi}(\frac{dB'_{z}}{dz}-ik_{x}B'_{z}),
\end{equation}
\begin{equation}\label{eq:4}
i \varphi E'+(E_{\rm 0}+P_{\rm 0})(ik_{\rm x}u'_{\rm x}+\frac{du'_{\rm z}}{dz})=0,
\end{equation}
\begin{equation}\label{eq:4a}
i\varphi B'_{x}=-B_{0}\frac{du'_{z}}{dz},
\end{equation}
%
%\begin{equation}\label{eq:4b}
%i\varphi B'_{y}=-iB_{0}k_{x}u'_{y},
%\end{equation}
%%
\begin{equation}\label{eq:4c}
i\varphi B'_{z}=-iB_{0}k_{x}u'_{z},
\end{equation}
where $\varphi = \omega + k_x U_0 $. Also, we have
\begin{equation}\label{eq:5}
p'_{\rm g}=c_{\rm s}^{2}\rho',
\end{equation}
\begin{equation}\label{eq:6}
e'_{\rm g}=\frac{p'{g}}{\gamma-1},
\end{equation}
\begin{equation}\label{eq:7}
e'_{\rm r}=3p'_{\rm r}.
\end{equation}
Having equations (\ref{eq:1}), (\ref{eq:5}), (\ref{eq:6}) and (\ref{eq:7}), then the linearized energy equation (\ref{eq:4}) simplifies to
\begin{equation}\label{eq:7a}
p'_{\rm r}= \xi p'_{\rm g},
\end{equation}
where
\begin{equation}
\xi = \frac{1}{3}(1+4\beta ),
\end{equation}
and  $\beta=p_{\rm 0r}/p_{\rm 0g}$ and we assume that the radiation pressure is smaller than the gas pressure which implies   $0 \leq \beta \leq 1$ and  $\frac{1}{3} \leq \xi \leq \frac{5}{3}$.

Considering equations (\ref{eq:2}) and (\ref{eq:5}), we can also rewrite equation (\ref{eq:1}) as
\begin{equation}
p'_{\rm g} = \left ( \frac{- \rho_0 i }{ 1+\xi } \right ) \frac{1}{\left [ k_{\rm x}^2 - \frac{\varphi^2 }{c_{\rm s}^2 \left (1+\xi \right )}\right ]} \frac{du'_{\rm z}}{dz}
\end{equation}
Substituting the above equation and equation (\ref{eq:7a}) and  equations (\ref{eq:4a}) and (\ref{eq:4c}) into equation (\ref{eq:3}), the following  differential equation for $u'_{\rm z}$ is obtained
\begin{equation}
\frac{d^{2}u'_{\rm z}}{dz^{2}}-q^{2} u'_{\rm z}=0,
\end{equation}
where
\begin{equation}
q^{2}=\bigg[1-\frac{B_{0}^{2}k_{x}^{2}}{4\pi\varphi^{2}\rho_{0}}\bigg]\bigg[{\big({k_{x}^{2}}-
\frac{\varphi^{2}}{c_{s}^{2}(1+\xi)})^{-1}-\frac{B_{0}^{2}}{4\pi\varphi^{2}\rho_{0}}\big)}\bigg]^{-1}.
\end{equation}

This is the main equation for analyzing KH instability, in which both the magnetic field and the radiation are considered. The general solutions for the upper and the lower layers are written as
\begin{eqnarray*}
\left\{
\begin{array}{rl}
u'_{\rm 1z}=C_{\rm 1}e^{-q_{\rm 1}z}+C'_{\rm 1}e^{+q_{\rm 1}z} &\qquad    z > 0\\
u'_{\rm 2z}=C_{\rm 2}e^{+q_{\rm 2}z}+C'_{\rm 2}e^{-q_{\rm 2}z} &\qquad    z < 0
\end{array} \right.
\end{eqnarray*}
where $C_{\rm 1}$, $C_{\rm 2}$, $C'_{\rm 1}$ and $C'_{\rm 2}$ are constants to be  determined based on the
boundary conditions. Here, the parameters $q_1$ and $q_2$ are corresponding to the upper and the lower regions, respectively. Without losing the generality of the problem, we assume both these parameters have positive real parts. Once we obtain  unstable modes, it will be checked if our assumption regarding to  $q_1$  and $q_2$ are satisfied.

One boundary condition is that the flow should not diverge at infinity, i.e. $u'_{\rm z}=0$ when $z \rightarrow \pm \infty$. Thus,
\begin{eqnarray*}
\left\{
\begin{array}{rl}
u'_{\rm 1z}=C_{\rm 1}e^{-q_{\rm 1}z} &\qquad    z > 0\\
u'_{\rm 2z}=C_{\rm 2}e^{+q_{\rm 2}z} &\qquad    z < 0
\end{array} \right.
\end{eqnarray*}
Another boundary condition is the continuity of the displacement
across the interface $z=0$. So, we obtain
\begin{equation}
\frac{C_{\rm1}}{\varphi_{\rm 1}}=\frac{C_{\rm2}}{\varphi_{\rm 2}}.
\end{equation}
 Finally, the third boundary condition is the continuity  of the total pressure at the interface.  Now, we can obtain the dispersion relation.  We assume the initial density and the sound speed at the upper layer are   $\rho_{1}$  and $c_{s1}$, respectively. Also, $\rho_{2}$ and $c_{s2}$ are the initial density and the sound speed at the lower layer. Therefore, the dispersion equation is obtained
\begin{eqnarray}\label{eq:8}
&&\alpha ({x+1})^{2}[1-\frac{1}{M_{1B}^{2}(x+1)^{2}}]^{\frac{1}{2}}\nonumber\\
&&\bigg[\frac{1}{1 -\frac{M_{1}^{2}(x+1)^{2}}{(1+\xi)}}-\frac{1}{M_{1B}^{2}(x+1)^{2}}\bigg]^{\frac{1}{2}}+\nonumber\\
&&({x-1})^{2}[1-\frac{1}{M_{2B}^{2}(x-1)^{2}}]^{\frac{1}{2}}\nonumber\\
&&\bigg[\frac{1}{1-\frac{M_{2}^{2}(x-1)^{2}}{1+\xi}}-\frac{1}{M_{2B}^{2}(x-1)^{2}}\bigg]^{\frac{1}{2}}=0
\end{eqnarray}
where $M_{1}$ and $M_2$ are Mach numbers in each layer, i.e. $M_{1} = {U_{0}}/{c_{s1}}$ and $M_{2} = {U_{0}}/{c_{s2}}$. Also, $x$ is the  dimensionless  growth rate which is defined as $x = \omega/k_{\rm x}U_{\rm 0}$ and the density contrast of the layers is denoted by  $\alpha = \rho_{1}/\rho_{2}$. Also, magnetic Mach number are defined as $M_{1B}={U_{0}}/{v_{1A}}$ and $M_{2B}={U_{0}}/{v_{2A}}$, where $v_{1A}$ and $v_{2A}$ are the Afven speed in each layer. If we expand the above equation, then a six order polynomial is obtained. We solve this equation numerically to find the unstable modes.

As another illustrative case, we consider two streaming layers where in one medium the radiation field is negligible but in the another medium the dynamical role of the radiation is considered.   If we don't have radiation in fluid 2, the final dispersion relation becomes
\begin{eqnarray}\label{eq:WR}
&&\alpha({x+1})^{2}[1-\frac{1}{M_{1B}^{2}(x+1)^{2}}]^{\frac{1}{2}}\nonumber\\
&&\bigg[\frac{1}{1 -\frac{M_{1}^{2}(x+1)^{2}}{(1+\xi)}}-\frac{1}{M_{1B}^{2}(x+1)^{2}}\bigg]^{\frac{1}{2}}+\nonumber\\
&&({x-1})^{2}[1-\frac{1}{M_{2B}^{2}(x-1)^{2}}]^{\frac{1}{2}}\nonumber\\
&&\bigg[\frac{1}{1-{M_{2}^{2}(x-1)^{2}}}-\frac{1}{M_{2B}^{2}(x-1)^{2}}\bigg]^{\frac{1}{2}}=0
\end{eqnarray}
We can also simplify the above relations by neglecting the magnetic field. When radiation field exists in both non-magnetized layers, one can simplify equation (\ref{eq:8}) as
\begin{equation}\label{eq:11}
\frac{\alpha({x+1})^{2}}{\sqrt{1-\frac{M_{1}^{2}(x+1)^{2}}{1+\xi}}}
+\frac{({x-1})^{2}}{\sqrt{1-\frac{M_{2}^{2}(x-1)^{2}}{1+\xi}}}=0.
\end{equation}
Also, for the non-magnetized case equation (\ref{eq:WR}) reduces to the following equation,
\begin{equation}\label{eq:12}
\frac{\alpha({x+1})^{2}}{\sqrt{1-\frac{M_{1}^{2}(x+1)^{2}}{1+\xi}}}
+\frac{({x-1})^{2}}{\sqrt{1-{M_{2}^{2}(x-1)^{2}}}}=0
\end{equation}

Equations (\ref{eq:8}), (\ref{eq:WR}), (\ref{eq:11}) and (\ref{eq:12}) are our dispersion relations for analyzing KH instability  under different conditions. In the next section, we study the unstable perturbations numerically.

\section{Analysis}
\begin{figure}[h]
%\vspace{-36pt}
%\epsfig{figure=f1.eps,angle=0,scale=0.5}
\includegraphics[width=8.3cm]{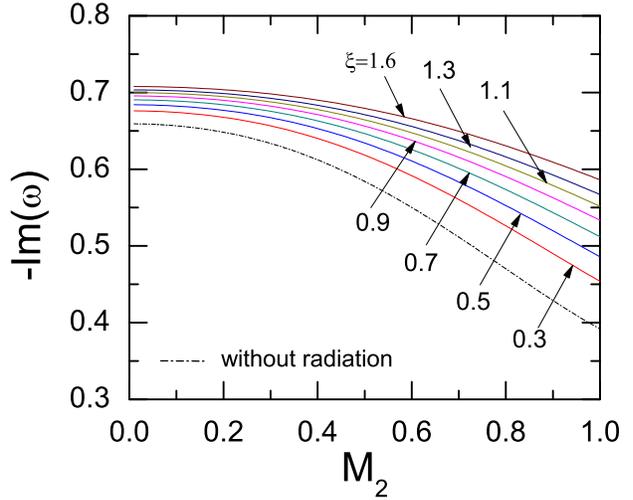}
\caption{Growth rate of the unstable perturbation versus Mach number for different values of the parameter $\xi$ when $M_{1}=0.7$ and $\alpha=2$. Each curve is labeled by the corresponding value of $\xi$ and the dashed-dot curve shows growth rate for a case without radiation.}
\label{fig:f1}
\end{figure}
\begin{figure}
%\vspace{-36pt}[ht]
%\epsfig{figure=f2.eps,angle=0,scale=1.1}
\includegraphics[width=7.7cm]{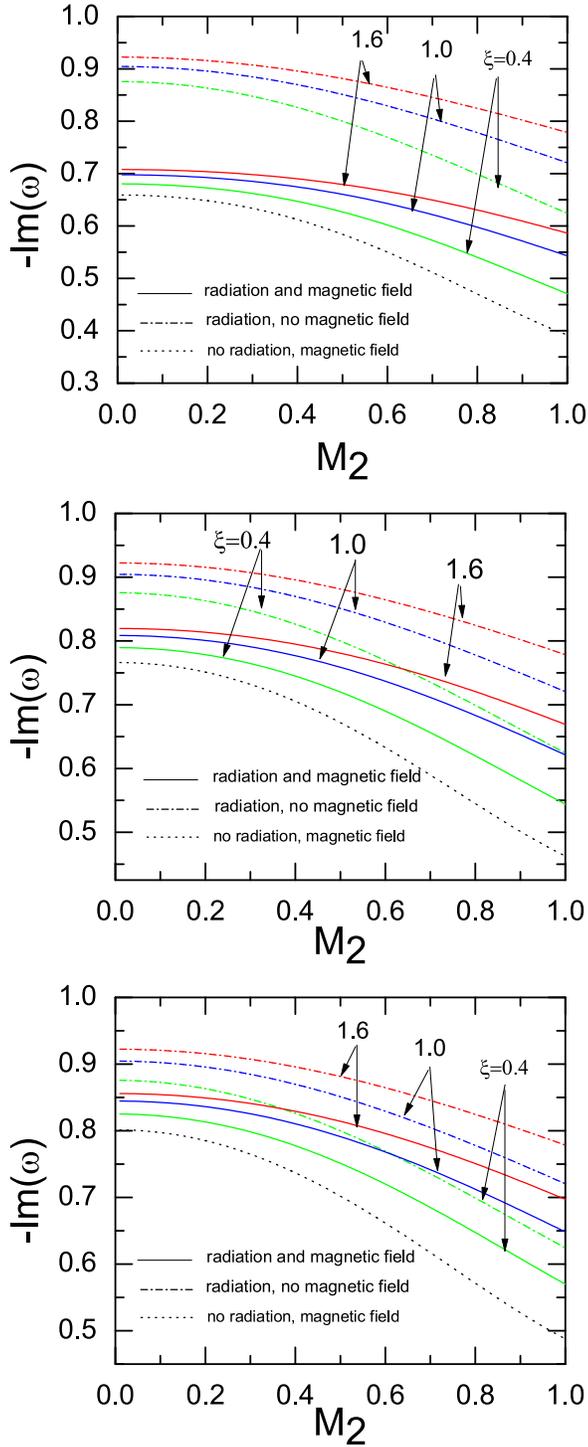}
\caption{Growth rate of the unstable perturbation versus Mach number for different values of the parameter $\xi$ for different  magnetic Mach numbers, i.e.  $M_{1B}=2$ (upper), $M_{1B}=3$ (middle) and $M_{1B}=4$ (bottom). Other input parameters are $M_{1}=0.7$ and $\alpha=2$. Each curve is labeled by the corresponding value of $\xi$ and the dashed-dot curve shows growth rate for a case without magnetic field and with the radiation. Also, dot curves show growth rate for a magnetized case but without the radiation.}
\label{fig:f2}
\end{figure}
\begin{figure}
%\vspace{-36pt}
%\epsfig{figure=f2'.eps,angle=0,scale=1.1}
\includegraphics[width=7.7cm]{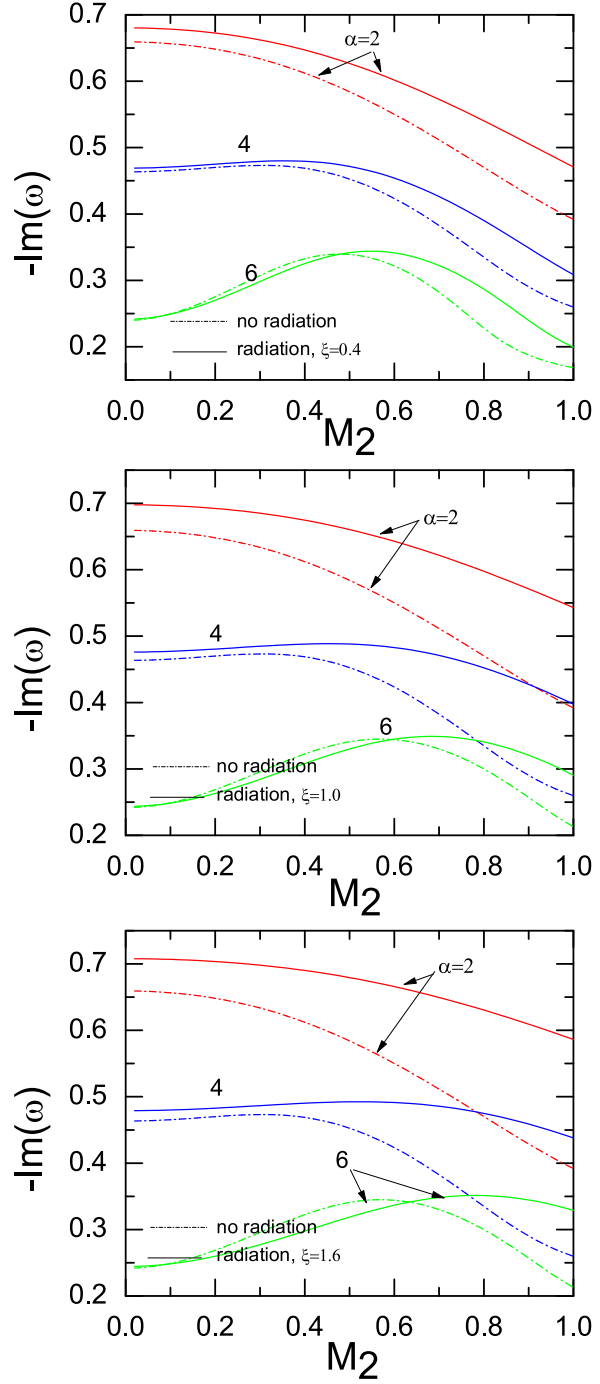}
\caption{Growth rate of the unstable perturbation versus Mach number for different values of the density contrast $\alpha$. Each curve is labeled by the corresponding value of $\alpha$ and for three values for the  radiation parameter, i.e. $\xi=0.4$ (upper),  $\xi=1.0$ (middle) and  $\xi=1.6$ (bottom). Dashed-dot curve shows growth rate for a case without the radiation. All plots are corresponding to  $M_{1}=0.7$ and $\alpha=2$.}
\label{fig:f3}
\end{figure}

First, we consider a non-magnetized system but with a radiation field. In this case, only one root of equation (\ref{eq:11}) is acceptable and corresponds to the instability. Figure \ref{fig:f1} shows growth rate of the unstable perturbation versus Mach number $M_{2}$ for different values of the radiation parameter $\xi$. Here, we have $M_{1}=0.7$ and $\alpha =2$ and the dashed-dot curve corresponds to a case without radiation. As we see  the curve without the radiation field has the lowest growth rate and  with increasing the strength of radiation field, the system becomes more unstable. In other words, dynamically important radiation field has a destabilizing effect on the growth of the unstable KH perturbations. However, the effect of the radiation is more significant at the larger Mach numbers.

In Figure \ref{fig:f2}, we explore the combined effects  of the magnetic field and the radiation.  In order to make easier comparison, growth rate corresponding to a magnetized case without radiation is shown by dot curve. Top, middle and bottom plots are corresponding to $M_{1B}=2,3,4$, respectively.   While radiation pressure tends to destabilize the system, we see that magnetic field has a stabilizing role. All growth rate curves shift downward because of the magnetic field. In particular, when the magnetic Mach number $M_{1B}$ is larger  magnetic field tends to compensate the destabilizing role of the radiation field.

In Figure \ref{fig:f3}, the effect of the density contrast $\alpha$ on the unstable KH perturbations is shown. In each plot, growth rates corresponding to the both cases with and without radiation are shown. The radiation parameter $\xi$ has three values, i.e. $\xi = 0.4$ (top), $1.0$ (middle) and  $1.6$ (bottom). Each curve is labeled by the corresponding value of $\alpha$. For a given radiation field, as the density contrast decreases,  the effect of the radiation becomes more significant. In particular, for our input parameters, when the density contrast is $\alpha =6$ the difference between the cases with and without radiation is negligible unless the Mach number becomes large where again the destabilizing role of the radiation field appears.

\section{Discussions}
We studied compressible KH instability in the presence of a dynamically important radiation field and the magnetic field.  Irrespective of the existence of the magnetic field, our results show that radiation field has a destabilizing effect on the unstable KH perturbations. One can simply define an effective pressure by adding thermal and radiation pressures. This will give us an effective sound speed. Thus, in  the presence of the radiation field the effective sound speed increases comparing to a case without radiation which then implies a less compressible system. In other words, dynamical role  of the radiation field leads to a reduction of  the compressibility of the system, and so,  KH perturbations have larger growth rates. Obviously, we do not expect to see the dynamical effect of the radiation field on the incompressible KH perturbations. On the other hand, magnetic field has a  stabilizing role. Thus, as we showed, there is always a competition between the radiation field and the magnetic field to control growth rate of the unstable KH perturbations. We also found when the density contrast of the layers is smaller, the dynamical role of the radiation is stronger. But as the density contrast increases, the effect of the radiation field becomes more significant  at the larger Mach numbers.

Our analysis is restricted to the linear regime and  so one can not apply the present results to the real  systems to obtain the mass and energy exchanges at the interface. But we expect stronger turbulence via KH instability at the nonlinear regime when dynamical role of the radiation becomes significant. One of the sources of the observed turbulence in ISM is the injection of mass and energy from the stellar outflows and jets. KH mechanism provides a way for these exchanges which would be amplified in the presence of the radiation field. Also, it is indeed likely that radiation pressure could have a dynamical effect in the boundaries of radiation driven bubbles where KH instability may occur. As a next step for the future work, numerical simulation of KH instability in the presence of the radiation is required to address these issues.

%\section*{Acknowledgments}
%We are grateful to the anonymous referee whose detailed and careful comments helped to improve the quality of this paper.

\nocite{*}
\bibliographystyle{spr-mp-nameyear-cnd}
%\bibliography{myref}
\bibliography{reference}

\end{document}